\documentstyle[12pt]{article}

\makeatletter
 
\@addtoreset{equation}{section}
\makeatother

\newcommand{\e}{\enskip}

\newcommand{\ts}{\hspace*}

\newcommand{\hs}{\hspace}
\newcommand{\vs}{\vspace}
\newcommand{\bc}{\begin{center}}
\newcommand{\ec}{\end{center}}
\newcommand{\f}{\frac}
\newcommand{\dps}{\displaystyle}
\newcommand{\pl}{\partial}

\newcommand{\xd}{\dot{x}}
\newcommand{\psid}{\dot{\psi}}
\newcommand{\ld}{\dot{\lambda}}
\newcommand{\Xd}{\dot{X}}
\newcommand{\Phid}{\dot{\Phi}}

\newcommand{\chid}{\dot{\chi}}
\newcommand{\Pid}{\dot{\Pi}}
\newcommand{\PXd}{\dot{P}^X} 

\newcommand{\Phat}{\hat{{\cal P}}}
\newcommand{\Pbhat}{\hat{{\sf P}}}
\newcommand{\Zp}{\hat{Z}^{(+)}}
\newcommand{\Zm}{\hat{Z}^{(-)}}
\newcommand{\Zpm}{\hat{Z}^{(\pm)}}
\newcommand{\Zmp}{\hat{Z}^{(\mp)}}
\newcommand{\Ap}{\hat{A}^{(+)}}
\newcommand{\Am}{\hat{A}^{(-)}}
\newcommand{\Apm}{\hat{A}^{(\pm)}}

\newcommand{\Bp}{\hat{B}^{(+)}}
\newcommand{\Bm}{\hat{B}^{(-)}}

\newcommand{\Bmp}{\hat{B}^{(\mp)}}
\newcommand{\calS}{{\cal S}}
\newcommand{\calC}{{\cal C}}

\newcommand{\pv}{p^{\Vert}}
\newcommand{\pn}{p^{\bot}}
\newcommand{\calZ}{{\cal Z}}

\begin{document}
\begin{center}
{\Large\bf $N=1$ Supersymmetric Quantum Mechanics  on a Curved Space} \vspace{16mm}\\
{\bf M. Nakamura}\footnote[2]{E-mail address: mnakamur@hamamatsu-u.ac.jp} and {\bf N. Okamoto}
\vs{4mm} \\
{\it Department of Physics, Tokoha-gakuen Hamamatsu University,Miyakoda-cho 1230, Hamamatsu-shi, Shizuoka 431-21, Japan} \vs{8mm}\\
{\bf H. Minowa}\footnote[8]{E-nail address: minowa@ci.thu.ac.jp}  \vs{4mm}\\
{\it Department of Physics, Teikyo Heisei University, Ichihara-shi, Chiba 290-01, Japan} \vspace{40mm}
\\
\end{center}
\begin{abstract}
The quantum mechanics of an $N=1$ supersymmetric dynamical system constrained to a hypersurface embedded in the higher dimensional Euclidean space is investigated by using the projection-operator method (POM) of constrained systems. It is shown that the {\it resulting} Hamiltonian obtained by the successive operations of projection operators contains the $\hbar^2$-contributing additional terms, which are completely missed when imposing constraints before the quantization. We derive the conditions the additional terms should satisfy when the $N=1$ supersymmetry holds in the {\it resulting} system, and present the geometrical interpretations of these additional terms. 
\end{abstract}
PACS 03.65.-w - Quantum theory;quantum mechanics \\
PACS 11.10.Ef - Lagrangian and Hamiltonian approach

\clearpage
\baselineskip 8mm

\section{Introduction}

\ts{12pt}The problem of the quantization of a dynamical system constrained to a submanifold embedded in the higher dimensional Euclidean space has been extensively investigated\cite {JK,C,HIM,OFK,INTT} as one of the quantum theories on a curved space. When quantizing such a system, one often faces the operator-ordering problem. The supersymmetry has then played the important role in the ordering problem\cite{DMPH,AFFR}. So, it is extremely interesting to extend the quantum mechanics on the curved space to the case that the system possesses a supersymmetry and to investigate whether or not the supersymmetry of the system holds when the constraints are imposed. In this paper, we investigate these problems within the canonical quantization formalism. For this purpose, we consider the $N=1$ supersymmetric model constrained to the hypersurface $\Sigma^{n-1}$ embedded in the $n$-dimentional Euclidean space $R^n$ within the framework of operator formulation with the sufficient generality, that is, $\Sigma^{n-1}$ does not depend on the specific geometrical structures.\\
\ts{12pt}Although such a system can be regarded as the constrained system with the second class constraints, there are two standard approaches to the quantization of constrained systems. The first approach\cite{Dirac,S} is to impose the constraints first and then to quantize on the reduced phase space (Approach A), and the second, inversely,  
first to quantize on the initial {\it flat} phase space, where the suitable set of canonically conjugate operators is well-defined, and then to impose the constraints as the {\it operator-equations} (Approach B). Then, there often occur the situations where the two are not equivalent. This discrepancy problem  has been extensively discussed until now\cite{DJT,W,L,J}. Then, it is shown that the  approach B involves the contributions, which are completely missed in the approach A\cite{DJT}, and it is pointed out that the approach B is more advantageous for the self-adjointness problem of unbounded operators\cite{J}. In the problem of quantization on the curved space, one of the algebraic formulations in the approach B has been recently proposed by Ohnuki and Kitakado\cite{OK}, where the so-called {\it induced} gauge potentials are derived.  In the second-class constrained systems, one of the approach B has been proposed by Batalin and Fradkin (BF)\cite{BF} as the basis of the quantization of constrained systems with the path integral formulation.  In the context of the canonical formalism of quantization, we have proposed the alternative method called the projection operator method (POM), which is shown to be equivalent to the BF method at the level of operator-algebra\cite{NM,NH}. In Appendix A, we briefly review the POM. \\
\ts{12pt}The problem of quantizing the dynamical system constrained to a submanifold embedded in the Euclidean space has been mostly considered in the approach A. Therefore, it is very interesting to investigate this problem in the approach B. We investigate the above-mentioned $N=1$ supersymmetric model by using the POM, and show that the the {\it resulting} Hamiltonian obtained through the successsive operations of a series of projection operators contains the $\hbar^2$-contributing additional terms, which are completely missed in the apporoach A and therefore differ from the so-called quantum potentials appearing in the quantization on curved spaces\cite{DW}.\\ 
\ts{12pt}Since we treat the fermionic operators together with the bosonic ones, we shall adopt the supercommutator as the commutator of operators $A$ and $B$\cite{DW2},
$$
[A\ ,\  B] = AB - (-1)^{\epsilon (A)\epsilon (B)}BA, \eqno{(1.1)}
$$
and the supersymmetrized product of these operators,
$$
\{A\ ,\  B\}_S = \frac12(AB + (-1)^{\epsilon (A)\epsilon (B)}BA), \eqno{(1.2)} 
$$
where $\epsilon (A)$ denotes the Grassmann parity of the operator $A$. Then, we introduce $\Ap$ and $\Am$ operating on an operator $O$ as follows, respectively:
$$
\Ap O = \{A, O\}_S,
\eqno{(1.3)}
$$
$$
\Am O = \f1{i\hbar}[A, O], 
\eqno{(1.4)}
$$
which satisfy the usefull algebraic relations
$$
\Ap\Bp O = (-1)^{\epsilon (A) \epsilon (B)}\Bp\Ap O +  \f14 [[A, B], O],
\eqno{(1.5a)}
$$
$$
\Am\Bm O = (-1)^{\epsilon (A) \epsilon (B)}\Bm\Am O + (\f1{i\hbar})^2[[A, B], O],
\eqno{(1.5b)}
$$
$$
\Apm\Bmp O = (-1)^{\epsilon(A) \epsilon (B)}\Bmp\Apm O + (\f1{i\hbar})\{[A, B], O\}_S.
\eqno{(1.5c)}
$$
\ts{12pt}This paper is organized as follows. In sec. 2, we present the model Lagrangian to provide the $N=1$ supersymmetric action on a superspace. We first quantize the system by the canonical quantization scheme, and then construct the set of constraint operators in such a manner as the consistency conditions for the time evolution of constraint operators hold.  In sec. 3, we derive the {\it resulting} constrained system, which describes to be constrained to the hypersurface $\Sigma^{n-1}$ embedded in  the Euclidean space $R^n$, by using the POM. It is shown that the {\it resulting} Hamiltonian contains the additional terms, which are classified into three types of the $\hbar^2$-contributing terms. In  sec. 4, we derive from the commutator algebra of the supercharge the supersymmetric Hamiltonian, which contains the $\hbar^2$-contributing term corresponding to one of the additional terms in the {\it resulting} Hamiltonian. Because of the difference of the {\it resulting} system does not always preserve the $N=1$ supersymmetry. Then, we prove that the {\it resulting} Hamiltonian has the $N=1$ supersymmetry when the other two types of additional terms in the Hamiltonian are commutable with the supercharge, and present the simple example of the sphere $S^{n-1}$ in $R^n$. In sec. 5, we develop the geometrical interpretations of the additional terms appearing in our constrained model. In sec. 6, the discussions and the concluding remarks are given.

\section{$N=1$ Supersymmetric Model}

\subsection{The description of model}

\ts{12pt}Let the $R^{n}$ be $n-$dimensional Euclidean space spanned by the Cartesian coordinates $x^i \e (i = 1, \cdots ,n)$, and $\Omega$, the space of a real Grassmann variable $\theta$. Further, let $\Phi^i \e (i = 1, \cdots ,n)$  be the real superfields defined on the superspace $R^n \times \Omega$ by 
$$
\Phi^i = x^i + i\theta \psi^i\hs{24pt} (i = 1, \cdots ,n),
\eqno{(2.1)}
$$
and $\Lambda$, the auxiliary superfield defined on $R \times \Omega$ by
$$
\Lambda = - \lambda + \theta X ,
\eqno{(2.2)}
$$
where $\psi^i$ and $\lambda$ are the real Grassmann variables, and $X$, the real  bosonic one. Then, we start with the $N=1$ supersymmetric Lagrangian 
$$
L = \f{i}{2}\int d\theta (\Phid^iD\Phi^i + \Lambda G(\Phi)),
\eqno{(2.3)}
$$ 
where $D$ is the covariant derivative $D = \pl_\theta - i\theta \pl_t$, which satisfies $D^2 = -i\pl_t$. In the Lagrangian (2.3), the constraint superfunction $G(\Phi)$ is defined by
$$
G(\Phi) = G(x) + i\theta \psi^iG_i(x),
\eqno{(2.4)}
$$
where
$$
G_{i\cdots j}(x) = \pl_i\cdots \pl_jG(x), \hs{36pt} \pl_i = \pl^i = \f{\pl}{\pl x^i}.
\eqno{(2.5)}
$$
The action $S = \dps{\int}dt L$ is then invariant under the supertransformation
$$
\delta\Phi^i = - \varepsilon{\cal Q}\Phi^i,\hs{36pt} \delta\Lambda = - \varepsilon{\cal Q}\Lambda,                                                                            
\eqno{(2.6)}
$$
where ${\cal Q} = \pl_\theta + i\theta \pl_t,$ which satisfies ${\cal Q}^2 = i\pl_t$ and $[D,\ {\cal Q}] = 0$, and $\varepsilon$ is a Grassmann real parameter.

\subsection{Canonical quantization and constraints}

\ts{12pt}Carrying out  the integration with respect to $\theta$ in the Lagrangian (2.3),  we obtain
$$
L = \f12\xd^i \xd^i + \f i2\psi^i\psid^i + i\lambda G_i(x)\psi^i + XG(x).
\eqno{(2.7)}
$$
We first quantize the system  by the canonical quantization scheme. The canonical momenta conjugate to the variables $x^i, \psi^i, \lambda$ and $X$ are given by
$$
p_i = \f{\pl L}{\pl \xd^i} = \xd^i,
\eqno{(2.8a)}
$$
$$
\Pi_i = \f{\pl L}{\pl \psid^i} = - \f i2\psi^i,\hs{24pt} \Pi_{\lambda} = \f{\pl L}{\pl \ld} = 0,\hs{24pt} P^X = \f{\pl L}{\pl \Xd} = 0,
\eqno{(2.8b)}
$$
respectively. Let us express the set of the {\it initial} canonically conjugate pairs $(x^i, p_i)$, $(\psi^i, \Pi_i)$, $(\lambda, \Pi_\lambda)$ and $(X, P^X)$  by $\calC^{(0)}$,
$$
\calC^{(0)} = \{(x^i, p_i), (\psi^i, \Pi_i), (\lambda, \Pi_\lambda), (X, P^X)\}.
\eqno{(2.9)}
$$
The canonical commutation relations of $\calC^{(0)}$ are defined by
$$
[x^i, p_j] = i\hbar\delta^i_j,\hs{12pt}[X, P^X] = i\hbar,\hs{12pt}[\psi^i, \Pi_j] = - i\hbar\delta^i_j,\hs{12pt}
[\lambda, \Pi_\lambda] = - i\hbar,
\eqno{(2.10)}
$$
and the others, zero. \\
\ts{12pt}Eqs. (2.8b) give rise to the primary constraints
$$
\chi_i = \Pi_i + (i/2)\psi^i = 0,\hs{12pt}\Pi_\lambda = 0,\hs{12pt}P^X = 0.
\eqno{(2.11)}
$$
Thus, we have the constraint operators $\chi_i,\e \Pi_\lambda$ and $P^X$. Then, the {\it primary} Hamiltonian $H_P$ is represented as 
$$
H_P = H_0 + H'.
\eqno{(2.12)}
$$
Here $H_0$ and $H'$ are given by
$$
H_0 = \f12p_ip_i,
\eqno{(2.13a)}
$$
$$
H' = - i\lambda\psi^iG_i(x) - XG(x) +\{\mu^i, \chi_i\}_S + \{\tau, \Pi_\lambda\}_S + \{u, P^X\}_S,
\eqno{(2.13b)}
$$
respectively, where $\mu^i,\e \tau$ and $u$ are the Lagrange multiplier operators corresponding to the unknown {\it velocity} operators $\psid^i,\e \ld$ and $\Xd$, respectively. \\
\ts{12pt}We next consider the consistency conditions for the time evolutions of primary constraints (2.11). For a constraint operator $K$, such a condition is given by $\dot{K} = (1/i\hbar)[K, H] = 0$, which, besides the case that $K$ commutes with the Hamiltonian $H$, produces a series of secondary constraints until the Lagrange multipliers are determined. \\
\ts{12pt}From the consistency conditions for the constraints (2.11), we obtain the secondary constraints
$$
\begin{array}{l}
\eta_1 = \psi^iG_i(x) = 0,\vs{12pt}\\
\eta_2 = \lambda g^2 + \{\psi^iG^j_i(x), p_j\}_S = 0,\vs{12pt}\\
h_1 = G(x) = 0,\vs{12pt}\\
h_2 = \{G^i(x), p_i\}_S = 0,\vs{12pt}\\
h_3 = g^2X + i\lambda\psi^iG^j_i(x)G_j(x) + \{\{G^{ij}(x), p_i\}_S, p_j\}_S = 0
\end{array}
\eqno{(2.14)}
$$
with
$$
g^2 = G^i(x)G_i(x),
\eqno{(2.15)}
$$ 
and  the Lagrange multiplier operators $\mu^i$, $\tau$ and $u$, which are given in Appendix B together with the derivation processes of constraints (2.14). Thus, we have the consistent set of constraint operators,
$$
\calS^{(0)} = \{\chi_i, \Pi_\lambda, \eta_1, \eta_2, P^X, h_1, h_2, h_3\},
\eqno{(2.16)}
$$
which are obviously second class. \\
\ts{12pt}Let $(\calC, H, \calS)$ be the quantum system defined by  the canonically conjugate set $\calC$, the Hamiltonian $H$ and the set of constraint operators $\calS$. Then, the above-obtained {\it initial} system is expressed as $(\calC^{(0)}, H_P, \calS^{(0)})$.

\section{Projection of Operators}

\subsection{Construction of projection operators}

\ts{12pt}Observing the structure of the commutator algebra of $\calS^{(0)}$, we find that $\calS^{(0)}$ is convenient to be classified into the following five subsets:
$$
\calS_\chi = \{\chi_i\},\hs{12pt}\calS_G = \{h_1 ,h_2\},\hs{12pt}\calS_X =\{h_3, P^X\},\hs{12pt}\calS_\lambda = \{\eta_2 , \Pi_\lambda\},\hs{12pt}\calS_\eta = \{\eta_1\}.
\eqno{(3.1)}
$$
Then, our task is to {\it reduce} $\calC^{(0)}$ in such a manner as the {\it reduced} canonical operators satisfy the constraints (2.11) and (2.14), and to represent the Hamiltonian (2.12) in terms of these {\it reduced} operators. Using the POM, we shall accomplish such {\it reductions} of operators through the successive operations of the projection operators corresponding to the subsets (3.1).\\
\ts{12pt}Following the POM, we first construct the {\it ACCS} of the subsets (3.1). They are given as follows:
$$
\begin{array}{l}
\calS_\chi \e :\e  z_i = \hbar^{-1/2}\chi_i,\vs{12pt}\\
\calS_G\e :\e \left\{ \begin{array}{l}
\xi_G = h_1,\\
\pi_G = \{g^{-2} , h_2\}_S = \{g^{-1}n^i, p_i\}_S,
\end{array}\right.\vs{12pt}\\
\calS_X\e :\e \left\{\begin{array}{l}
\xi_X = \{g^{-2} , h_3\}_S + \dps{\f14}g^{-2}[g^{-2}, [g^2, h_3]],\\
\pi_X = P^X,
\end{array}\right.\vs{12pt}\\
\calS_\lambda\e :\e \left\{\begin{array}{l}
\xi_\lambda = - \lambda - \{g^{-2}\psi^iG^j_i(x) , p_j\}_S,\\
\pi_\lambda = \Pi_\lambda ,
\end{array}\right.
\end{array}
\eqno{(3.2)}
$$
where 
$$
n^i = n_i = g^{-1}G_i(x),
\eqno{(3.3)}
$$
which is the operator corresponding to the vector field normal to the hypersurface $\Sigma^{n-1}$, and satisfies
$$
n^in_i = 1.
\eqno{(3.4)}
$$
Here, the additional term $\dps{\f14}g^{-2}[g^{-2}, [g^2,h_3]]$ in the representation of $\xi_X$ has been needed for $\Phat_X$ to satisfy the {\it projection condition} $\Phat_Xh_3(\calC) = h_3(\Phat_X\calC) = 0$. 
Then, the corresponding projection operators are defined by (A.6). Let $\Phat_\chi,\e \Phat_G,\e \Phat_X$ and $\Phat_\lambda$ be the projection operators for the subsets $\calS_\chi,\e \calS_G,\e \calS_X $and $\calS_\lambda$, respectively. Under the operation of $\Phat_\chi$ , the commutator of $\eta_1$ becomes
$$
[\eta_1 , \eta_1] = \hbar g^2.
\eqno{(3.5)}
$$
Thus, the {\it ACCS} for $\calS_\eta$ is given by
$$
\hs{-136pt}\calS_\eta\e :\e z_\eta = \f{1}{\sqrt{\hbar}g}\eta_1.
\eqno{(3.6)}
$$
Let $\Phat_\eta$ be the projection operator constructed with the {\it ACCS} (3.6), which is defined under the operation of $\Phat_\chi$.

\subsection{Successive operations of projection operators}

\ts{12pt}Let $\Pbhat$ be a product of the projection operators $\Phat_\chi$, $\Phat_G$, $\Phat_X$, $\Phat_\lambda$ and $\Phat_\eta$, for example, $\Pbhat  = \Phat_X\Phat_\lambda\Phat_G\Phat_\eta\Phat_\chi$. We shall call $\Pbhat$ the successive projection. The projection operators in $\Pbhat$ are not always commutable with each other (see Appendix A.4). In such a case, $\Pbhat$ becomes not to be projective, and the operation of $\Pbhat$ depends on the order of the successive operations of the projection operators in $\Pbhat$ ({\it projection-ordering}). Sequentially using the {\it commutator-formula} (A.12a) and the {\it product-formula} (A.12b), then, we obtain the following results.\\
(1) {\bf The projections of the {\it initial} canonically conjugate set $\calC^{(0)}$}\\
\ts{12pt}Taking account of the structure of the commutators of $\calC^{(0)}$ with the {\it ACCS} (3.2) and (3.6), we find that the projection of $\calC^{(0)}$ depends on no {\it  projection-ordering} of projection operators. Under the operation of any $\Pbhat$, therefore, $\calC^{(0)}$ is {\it reduced} into the set of projected operators, $\calC^{(R)}$, as follows:
$$
\calC^{(R)} = \Pbhat\calC^{(0)} = \{x^i , p_i, \psi^i\}\hs{36pt}(i = 1,\cdots ,n),
\eqno{(3.7)}
$$
where $x^i = \Pbhat x^i,\e p_i = \Pbhat p_i$ and $\psi^i = \Pbhat \psi^i$.  
Then, the operators in $\calC^{(R)}$ satisfy the operator-constraint conditions 
$$
G(x) = 0,\hs{12pt}\{n^i(x) , p_i\}_S = 0,\hs{12pt}n_i(x)\psi^i = 0,
\eqno{(3.8)}
$$
the set of which we express as $\calS^{(R)}$, and the commutator algebra of  $\calC^{(R)}$ is given as follows:
$$
\begin{array}{ll}
&[x^i , p_j] = i\hbar W^i_j,\vs{12pt}\\
&\begin{array}{ll}
\hs{-4pt}[p_i , p_j] = & i\hbar\{n_j\partial^kn_i - n_i\partial^kn_j , p_k\}_S \vs{8pt}\\
              & -\hbar (\delta^k_i\delta^m_j-\delta^k_in_jn^m-\delta^m_jn_in^k)G_{kl:mn}\{\psi^l , \psi^n\}_S,
\end{array}\vs{12pt}\\
&[\psi^i , \psi^j] = \hbar W^{ij},\vs{12pt}\\
&[\psi^i , p_j]  = -i\hbar n^iW^k_j\partial_kn_l\psi^l,
\end{array}
\eqno{(3.9)}
$$
and the others are zero, where
$$
G_{kl:mn} =g^{-2}G_{kl}(x)G_{mn}(x),
\eqno{(3.10)}
$$
$$
W_{ij} = \delta_{ij} - n_in_j.
\eqno{(3.11)}
$$It should be noted that the commutator algebra (3.9) is just equivalent to the commutator algebra constructed from the corresponding Dirac brackets.\\ 
\ts{12pt}The remaining operators in $\calC^{(0)}$ are expressed in terms of the projected operators in $\calC^{(R)}$ as follows:
$$
\begin{array}{l}
\Pi_i = -\f{i}2\psi^i,\vs{12pt}\\
\lambda = -\{g^{-2}, \{\psi^iG^j_i(x), p_j\}_S\}_S,\vs{12pt}\\
X = \{g^{-2}, X_0\}_S + \f14[g^2, [g^{-2}, \{g^{-2} , X_0\}_S]]
\end{array}
\eqno{(3.12a)}
$$
with
$$
X_0 = -i\Pbhat\{\lambda, \psi^iG^j_i(x)G_j(x)\}_S - \Pbhat\{\{G^{ij}, p_i\}_S, p_j\}_S,
\eqno{(3.12b)}
$$
and $\Pi_\lambda\e =\e P^X\e =\e 0$.\\
(2) {\bf The projection of the {\it primary} Hamiltonian $H_P$}\\
\ts{12pt} Let $H_R$ be the projection of the {\it primary} Hamiltonian  $H_P$ defined by (2.12), 
$$
H_R = \Pbhat H_P.
\eqno{(3.13)}
$$
We call the quantum system $(\calC^{(R)}, H_R, \calS^{(R)})$ the {\it resulting} system. Sequentially using the {\it commutator-formula} (A.12a) and the {\it product-formula} (A.12b) with the troublesome but straightforward calculations, then, we find that $H_R$ contains the various kinds of the additional terms depending on the {\it projection-ordering} of $\Pbhat$. These terms are completely missed in the qunatization scheme due to the Approach A, and are interpreted as the quantum corrections caused by the {\it reductions} of operators. In order to represent these additional terms more clearly, let us introduce the following notations:
$$
\begin{array}{l}
G^{WW}(x) = G_{ij:kl}W^{ik}W^{jl},\vs{12pt}\\
G^{WN}(x) = G^{NW}(x) = G_{ij:kl}W^{ik}N^{jl},\vs{12pt}\\
G^{NN}(x) = G_{ij:kl}N^{ik}N^{jl},
\end{array}
\eqno{(3.14)}
$$
where $N^{ij}$ is defined by
$$
N^{ij} = n^i(x)n^j(x).
\eqno{(3.15)}
$$
The operators $W^{ij}$ and $N^{ij}$ satisfy the following available relations:
$$
W^{ij}n_j = n^iW_{ij} = 0,
\eqno{(3.16a)}
$$
$$
N^{ij}n_j = n^i,\hs{12pt}n^iN_{ij} = n_j,
\eqno{(3.16b)}
$$
$$
W_{ik}W^{kj} = W_i^j,\hs{12pt}N_{ik}N^{kj} = N_i^j,
\eqno{(3.17)}
$$ 
$$
W^{ij}N_{jk} = N^{ij}W_{jk} = 0,
\eqno{(3.18)}
$$
$$
W^{ij} + N^{ij} = \delta^{ij}.
\eqno{(3.19)}
$$
\ts{12pt}Consider first the projection of $H_0$ given by (2.13a). Taking account of the commutators of $p_i$ with the {\it ACCS}, we find that the projection of $H_0$ depends on only the {\it projection-ordering} of $\Phat_G$ and $\Phat_\eta$. So, the successive projections are classified into the following two types: One is $\Pbhat$ in which the operation of $\Phat_G$ is carried out before the operation of $\Phat_\eta$, done, which we express by $\Pbhat^I$, and the other, the reverse of $\Pbhat^I$ with respect to the operations of $\Phat_G$ and $\Phat_\eta$, which we express by $\Pbhat^{II}$. Then, we obtain the projection of $H_0$, which we express by $H^R_0$, as follows: 
$$
H_0^R = \Pbhat H_0 = \f12p_ip_i + H_0^Q(x),
\eqno{(3.20a)}
$$
where the additional term $H_0^Q(x)$ is given by
$$
H_0^Q(x) = \left\{\begin{array}{l}
H_I^Q(x) = -\dps{\f{\hbar^2}8}(G^{WW}(x) + G^{WN}(x) - G^{NN}(x))\hs{24pt}(\Pbhat = \Pbhat^I),\vs{12pt}\\
H_{II}^Q(x) = -\dps{\f{\hbar^2}8}(G^{WW}(x) - G^{NN}(x))\hs{78pt}(\Pbhat = \Pbhat^{II}),
\end{array}\right.
\eqno{(3.20b)}
$$
with $(x^i , p_i) \in \calC^{(R)}$. \\
\ts{12pt}We next consider the projection of $H'$, which depends on the {\it  projection-ordering} of all the projection operators $\Phat_\chi$, $\Phat_G$, $\Phat_X$, $\Phat_\lambda$ and $\Phat_\eta$. Sequentially using the formulas (A.12a) and (A.12b) with the rather tedious calculations, we find that the projection $\Pbhat H'$ consists of only the additional terms and takes such a form as
$$
\Pbhat H' = \f{\hbar^2}{4}(\alpha G^{WW}(x) + \beta G^{WN}(x) + \gamma G^{NN}(x)),
\eqno{(3.21)}
$$
where $\alpha$, $\beta$ and $\gamma$ are some integers. Consider,  then, the seccessive projections $\Pbhat_S$, which make the projection of $H'$ vanishing, {\it i.e.} $\alpha = \beta = \gamma = 0$ in (3.21). Such projections are realized by the following two: 
$$
\Pbhat_S = \left\{\begin{array}{l}
\Pbhat^I_S = \Phat_X\Phat_\lambda\Phat_\eta\Phat_G\Phat_\chi,\vs{12pt}\\
\Pbhat^{II}_S = \Phat_X\Phat_\lambda\Phat_G\Phat_\eta\Phat_\chi.
\end{array}\right.
\eqno{(3.22)}
$$
Then, the operations of $\Pbhat^I_S$ and $\Pbhat^{II}_S$ on $H_P$ become equivalent to $\Pbhat^IH_0$ and $\Pbhat^{II}H_0$, 
respectively. We thus obtain the {\it resulting} Hamiltonian
$$
H_S^R = \Pbhat_SH_P = \left\{\begin{array}{l}
\dps{\f12}p_ip_i + H_I^Q(x)\hs{24pt}(\Pbhat_S = \Pbhat_S^I),\vs{12pt}\\
\dps{\f12}p_ip_i + H_{II}^Q(x)\hs{24pt}(\Pbhat_S = \Pbhat_S^{II}).
\end{array}\right. 
\eqno{(3.23)}
$$
The {\it resulting} Hamiltonian $H_S^R$ contains no additional terms due to the auxiliary degrees of freedom. So, we call the projections $\Pbhat_S$ the {\it standard} projections, and $H_S^R$, the {\it standard} Hamiltonian. 

\section{Supersymmetry and Quantum Corrections}

\subsection{Commutator algebra of supercharge}

\ts{12pt}The supertransformation (2.6) is represented in terms of components as 
$$
\begin{array}{l}
\delta x^i = -i\varepsilon\psi^i,\vs{12pt}\\
\delta \psi^i = \varepsilon \xd^i,\vs{12pt}\\
\delta\lambda = -\varepsilon X,\vs{12pt}\\
\delta X =-i\varepsilon \ld.
\end{array}
\eqno{(4.1)}
$$ 
Due to Noether's theorem, the supercharge $Q$ generating (4.1) is given by
$$
Q = \xd^i\psi^i - \lambda G(x),
\eqno{(4.2)}
$$
which is quantized in the {\it initial} system  $(\calC^{(0)},H_P,\calS^{(0)})$.  Let $Q^{(0)}$ be the operator corresponding to $Q$ in $(\calC^{(0)},H_P,\calS^{(0)})$. Then, $Q^{(0)}$ becomes
$$
Q^{(0)} = \psi^ip_i - \lambda G(x).
\eqno{(4.3)}
$$
\ts{12pt}We next consider the projection of $Q^{(0)}$, which we express by $Q^{(R)}$, and the commutator algebra of $Q^{(R)}$. Taking account of the linearity of $Q^{(0)}$ with respect to $\psi^i$ and $p_i$, we see that $Q^{(R)}$ depends on no {\it projection-ordering}. For any $\Pbhat$, therefore, we obtain 
$$
Q^{(R)} = \Pbhat Q^{(0)} = \{\psi^i , p_i\}_S.
\eqno{(4.4)}
$$
The commutators of $Q^{(R)}$ with $\calC^{(R)}$ are calculated by using the commutator algebra (3.9) as follows: 
$$
\begin{array}{ll}
&[x^i , Q^{(R)}] = i\hbar\psi^i,\vs{12pt}\\
&[\psi^i , Q^{(R)}] = \hbar p_i,\vs{12pt}\\
&[p_i , Q^{(R)}] = -i\hbar \{n_i\psi^j\partial^kn_j , p_k\}_S.
\end{array}
\eqno{(4.5)}
$$
Then, the the commutator $[Q^{(R)}, Q^{(R)}]$ becomes 
$$
[Q^{(R)}, Q^{(R)}] = \hbar p_ip_i -i\hbar \{\psi^i, \{p_j, n_i\psi^k\partial^jn_k\}_S\}_S.
\eqno{(4.6)}
$$
Using the formula (1.5a) and the operator-constraints (3.8), the  double-symmetrized product in (4.6) is calculated to be 
$$
\{\psi^i, \{p_j, n_i\psi^k\partial^jn_k\}_S\}_S = -\f{i\hbar^2}4G^{WW}(x).
\eqno{(4.7)}
$$
We thus obtain 
$$
[Q^{(R)}, Q^{(R)}] = \hbar p_ip_i - \f{\hbar^3}{4}G^{WW}(x).
\eqno{(4.8)}
$$

\subsection{Quantum corrections in Hamiltonian}

\ts{12pt}In the $N=1$ supersymmetric quantum mechanics, the supercharge $Q$ obeys the supersymmetric algebra
$$
[Q, Q] = 2\hbar H,
\eqno{(4.9)}
$$
where $H$ is the supersymmetric Hamiltonian, which obviously commutes with $Q$, and therefore $H$ is invariant under the supersymmetric transformation corresponding to (4.1). Then, the problem is whether or not there exists $\Pbhat$ satisfying that $ \Pbhat H_P$ is commutable with the supercharge $\Pbhat Q$ in  the {\it resulting} system. \\
\ts{12pt}From (4.8) and (4.9), the supersymmetric Hamiltonian $H^{SUSY}$ associated with $Q^{(R)}$ is obtained  as follows:
$$
\begin{array}{lcl}
H^{SUSY} &=& \dps{\f1{2\hbar}}[Q^{(R)}, Q^{(R)}] \vs{12pt}\\
&=& \dps{\f12}p_ip_i + H^{SUSY}_Q, 
\end{array}
\eqno{(4.10a)}
$$
where $H^{SUSY}_Q$ is the additional term given by
$$
H^{SUSY}_Q = -\f{\hbar^2}8G^{WW}(x),
\eqno{(4.10b)}
$$ 
which is regarded as the quantum correction caused by the noncommutativity between $p_i$ and $\psi^i$ in $(\calC^{(R)}, H_R,\calS^{(R)})$. Then, $H^{SUSY}$ satisfies 
$$
[H^{SUSY}, Q^{(R)}] = 0.
\eqno{(4.11)}
$$
In order to estimate the commutator of $H_R$ with $Q^{(R)}$,  let us compare $H^{SUSY}$ with $H_R$. We first consider $H_0^R$. From (3.20) and (4.10), we obtain
$$
H_0^R = H^{SUSY} + \triangle H_0^Q(x),
\eqno{(4.12a)}
$$
where $\triangle H_0^Q$ is the discrepancy of $H_0^R$ from $H^{SUSY}$ given by
$$
\triangle H_0^Q = \left\{\begin{array}{l}
\triangle H_I^Q = \dps{\f{\hbar^2}8}(- G^{WN}(x) +  G^{NN}(x))\hs{24pt}(\Pbhat = \Pbhat^I),\vs{12pt}\\
\triangle H_{II}^Q = \dps{\f{\hbar^2}8}G^{NN}(x)\hs{102pt}(\Pbhat = \Pbhat^{II}).
\end{array}\right.
\eqno{(4.12b)}
$$
Since $\Pbhat H'$ consists of only the additional terms, then, $H_R$ can be rewritten from (3.21) in such a form as 
$$
H_R = H^{SUSY} + \triangle H^Q,
\eqno{(4.13a)}
$$
where $\triangle H^Q$ is the discrepancy of $H_R$ from $H^{SUSY}$, which is given as follows:
$$
\begin{array}{lcl}
\triangle H^Q &=& \triangle H_0^Q + \Pbhat H'\vs{12pt}\\
&=& \left\{\begin{array}{l}
\dps{\f{\hbar^2}8}(2\alpha G^{WW}(x) + (2\beta-1)G^{WN}(x) + (2\gamma + 1)G^{NN}(x)) \hs{12pt}(\Pbhat = \Pbhat^I),\vs{12pt}\\
\dps{\f{\hbar^2}8}(2\alpha 'G^{WW}(x) + 2\beta 'G^{WN}(x) + (2\gamma '+1)G^{NN}(x))\hs{36pt}(\Pbhat = \Pbhat^{II})
\end{array}\right.
\end{array}
\eqno{(4.13b)}
$$
\vs{6pt}
with some integers $\alpha$, $\alpha'$, $\beta$, $\beta'$, $\gamma$ and $\gamma'$. For the {\it standard} projections, then, the {\it standard} Hamiltonian becomes 
$$
H_S^R = H^{SUSY}+ \triangle H_S^Q,
\eqno{(4.14a)}
$$
where
$$
\triangle H_S^Q = \left\{\begin{array}{l}
\triangle H_I^Q\hs{24pt}(\Pbhat_S = \Pbhat_S^I),\vs{12pt}\\
\triangle H_{II}^Q\hs{24pt}(\Pbhat_S = \Pbhat_S^{II}).
\end{array}\right.
\eqno{(4.14b)}
$$
It is obvious that the factors $2\beta-1$, $2\gamma+1$ and $2\gamma'+1$ in (4.13b) can never vanish. From (4.11), (4.13) and (4.14), we find that there exist {\it no} sucessive projections, which eliminate the $G^{WN}(x)$-term and the $G^{NN}(x)$-term in the discrepancy $\triangle H^Q$, and that the {\it resulting} Hamiltonian $H_R$ commutes with $Q^{(R)}$ when $\triangle H^Q$ commuting with $Q^{(R)}$. Then, we obtain the following results:\\
(1) For all the successive projections $\Pbhat$, there exist {\it no resulting} Hamiltonians equivalent to the supersymmetric Hamiltonian,
$$
H_R = \Pbhat H_P \neq H^{SUSY}.
\eqno{(4.15)}
$$
Therefore, the $N=1$ supersymmetry the classical system possesses does not always hold in the {\it resulting} system.\\
(2) Let $(\calC^{(R)}, H_R, \calS^{(R)})$ be the {\it resulting} system {\it reduced} by the successive projection $\Pbhat$. If the discrepancy $\triangle H^Q$ commutes with the supercharge $Q^{(R)}$, then, $\Pbhat$ conserves the $N=1$ supersymmetry, that is, $H_R$ is invariant under the supertransformation with $Q^{(R)}$ in $(\calC^{(R)}, H_R, \calS^{(R)})$.\\ 
(3) For the {\it standard} projection $\Pbhat_S^{II}$, the discrepancy $\triangle H_S^Q$ contains only $G^{NN}(x)$-term. Therefore, $\Pbhat_S^{II}$ conserves the $N=1$ supersymmetry if $G^{NN}(x)$ commutes with the supercharge $Q^{(R)}$.\\
\ts{12pt}It should be noticed that this also occurs in the case of the Approach A with the Dirac bracket quantization, since, although the commutator algebra constructed from the Dirac brackets is equivalent to (3.9), the Hamilonian contains no additional terms. In the context of the Approach A, further, the problem of the symmetry breaking in various $N$ supersymmetric systems on manifolds has been investigated by Claudson and Halpern\cite{CH}   

\subsection{Simple example}

\ts{12pt}In order to illustrate the above-obtained results, let us consider the sphere $S^{n-1}$ as the hypersurface $\Sigma^{n-1}$ embedded in $R^n$. Then, $G(x)$ is given by
$$
G(x) = x^ix_i - R^2
\eqno{(4.16)}
$$ 
with a constant $R$. From (3.14), $G^{WW}(x)$, $G^{WN}(x)$ and $G^{NN}(x)$ are calculated to become
$$
G^{WW}(x) = \dps{\f{n-1}{R^2}},\hs{12pt}
G^{WN}(x) = 0,\hs{12pt}
G^{NN}(x) = \dps{\f1{R^2}}.
\eqno{(4.17)}
$$ 
For any successive projection $\Pbhat$, then, the discrepancy  $\triangle H^Q$ is given in the following form:
$$
\triangle H^Q = \f{\hbar^2}{8R^2}(2An + 2B + 1),
\eqno{(4.18)}
$$
where $A$ and $B$ some integers. The discrepancy (4.18) obviously commutes with $Q^{(R)}$. Thus, we see that the {\it resulting} Hamiltonian $H_R$ is invariant under the supertransformation with $Q^{(R)}$. 

\section{Interpretation of Quantum corrections}
  
\ts{12pt}The additional terms appearing in $H_R$ and $H^{SUSY}$ are seem to take the form similar to the $\hbar^2$-contribution terms arising through the quantization on a curved space with the Approach A, which  depend on the geometrical structures of the curved space. In our approach, however, the quantizaion is carried out on the unconstrained {\it flat} phase space, on which the suitable canonically conjugate set of operators is  well-defined, and the additional terms are caused by the noncommutativity of the {\it ACCS} with the operators of the system (see (A.13) and (A.14)). So, we shall attempt to develop the geometrical interpretation of these additional terms with the operator formalism. For this purpose, it is sufficient to concentrate our attensions into $H_S^R$, since the disappearance of the $N=1$ supersymmetry is caused essentially by the additional terms containing $G^{WN}(x)$ and $G^{NN}(x)$.\\
\ts{12pt} Let us introduce the decomposition of the operator in the {\it initial} system $(\calC^{(0)}, H_P, \calS^{(0)})$ into the component tangential to $\Sigma^{n-1}$ and the component normal to that. For an operator $f_i \in (\calC^{(0)}, H_P, \calS^{(0)})\e (i=1,\cdots ,n)$, the tangential component $f_i^{\parallel}$ and the normal component $f_i^{\perp}$ are defined in such a manner as they satisfy 
$$
\begin{array}{l}
\{n^i, f_i^{\parallel}\}_S = 0,\vs{12pt}\\
\{n^i, f_i^{\perp}\}_S = f^{\perp},
\end{array}
\eqno{(5.1)}
$$
respectively, where $f^{\perp}$ is given by
$$
f^{\perp} = \{n^i, f_i\}_S.
\eqno{(5.2)}
$$
When $f_i = f_i(x) = f^i(x)\e (i = 1,\cdots ,n)$, we readily obtain the decomposition
$$
f_i(x) = f_i^{\parallel}(x) + f_i^{\perp}(x),
\eqno{(5.3a)}
$$
where $f_i^{\parallel}(x)$ and $f_i^{\perp}(x)$ are defined by
$$
\begin{array}{l}
f_i^{\parallel}(x) = W_i^jf_j(x),\vs{12pt}\\
f_i^{\perp}(x) = N_i^jf_j(x),
\end{array}
\eqno{(5.3b)}
$$
respectively. They obviously satisfy the conditions (5.1) from (3.16) as
$$
\begin{array}{l}
\{n^i, f_i^{\parallel}\} =n^iW_i^jf_j(x)=0,\vs{12pt}\\
\{n^i, f_i^{\perp}\} =n^iN_i^jf_j(x)=f^{\perp}.
\end{array}
\eqno{(5.4)}
$$
The decomposition (5.3) is naturally extended to such an operator as $f_{ijk\cdots}(x)$. The tangential component and the normal one with respect to an index $k$ in $f_{ijk\cdots}(x)$ can be defined by
$$ 
\begin{array}{l}
W_k^sf_{ijs\cdots}(x),\vs{12pt}\\
N_k^sf_{ijs\cdots}(x),
\end{array}
\eqno{(5.5)}
$$
respectively. From (3.17), then, we see that $G^{WW}(x)$,  $G^{NN}(x)$ and $G^{WN}(x)$ consist of the tangential components, the normal ones and both the components, in  $G_{ij:kl}$, respectively.\\  
\ts{12pt}The decomposition of $p_i$ can be easily realized in the following way: Let us define 
$$
\begin{array}{l}
\pv_i = \{W_i^j, p_j\}_S,\vs{12pt}\\
\pn_i = \{N_i^j, p_j\}_S.
\end{array}
\eqno{(5.6)}
$$
These are readily shown to satisfy the conditions (5.1) by using (1.5a) and (3.16) as follows:
$$
\begin{array}{l}
\{n^i, \pv_i\}_S = \{n^iW_i^j, p_j\}_S = 0,\vs{12pt}\\
\{n^i, \pn_i\}_S = \{n^iN_i^j, p_j\}_S = \pn.
\end{array}
\eqno{(5.7)}
$$
Using (3.16), (3.17), (3.18) and (1.5a), we obtain
$$
\begin{array}{l}
\{W_i^j, \pv_j\}_S = \pv_i,\hs{12pt}\{N_i^j, \pv_j\}_S = 0,\vs{12pt}\\
\{W_i^j, \pn_j\}_S = 0,\hs{12pt}\{N_i^j, \pn_j\}_S = \pn_i.
\end{array}
\eqno{(5.8)}     
$$
From (3.19) and (5.6), thus, $p_i$ is decomposed as 
$$
p_i = \pv_i + \pn_i.
\eqno{(5.9)}
$$
Then, $H_0$ can be rewritten by using (5.9) in such a form as 
$$
H_0 = H_0^{(\parallel ,\parallel)} + H_0^{(\parallel ,\perp)} + H_0^{(\perp ,\perp)},
\eqno{(5.10a)}
$$
where $H_0^{(\parallel ,\parallel)}$, $H_0^{(\parallel ,\perp)}$ and $H_0^{(\perp ,\perp)}$ are given by
$$
H_0^{(\parallel ,\parallel)} = \dps{\f12}\pv_i\pv_i,\hs{24pt}H_0^{(\parallel ,\perp)} = \{\pv_i, \pn_i\}_S,\hs{24pt}H_0^{(\perp ,\perp)} = \dps{\f12}\pn_i\pn_i,
\eqno{(5.10b)}
$$
respectively.\\
\ts{12pt}Consider next the projections of decomposed operators in $(\calC^{(0)}, H_P, \calS^{(0)})$ into $(\calC^{(R)}, H_R, \calS^{(R)})$. Sequentially using the formulas (A.12a) and (A.12b), we readily obtain 
$$
\begin{array}{l}
\Pbhat \pv_i = \pv_i,\vs{12pt}\\
\Pbhat \pn_i = 0.
\end{array}
\eqno{(5.11)}
$$
Then, we call $\pv_i$ the {\it physical} component of $p_i$, and $\pn_i$, the {\it unphysical} one, under the operator-constraint conditions (3.8). In the projection of $f_i(x)$, on the other hand, there remains the projection of the normal component $f_i^{\perp}(x)$ together with that of the tangential component $f_i^{\parallel}(x)$ in the form-invariant manner as follows:
$$
\begin{array}{l}
\Pbhat f_i^{\parallel}(x) = f_i^{\parallel}(x),\vs{12pt}\\
\Pbhat f_i^{\perp}(x) = f_i^{\perp}(x).
\end{array}
\eqno{(5.12)}
$$
These projections satisfy the relations  
$$
\{f^{\parallel i}(x) , \pv_i\}_S = \{f^i(x), \pv_i\}_S,
\eqno{(5.13a)}
$$
$$
\{f^{\perp i}(x) , \pv_i\}_S = 0
\eqno{(5.13b)}
$$
from (5.8), that is, the tangential component $f^{\parallel}_i(x)$ is parallel to $\pv_i$ and the normal component $f^{\perp}_i(x)$, orthogonal to $\pv_i$ in $(\calC^{(R)} , H_R, \calS^{(R)})$. So, let us also call $f_i^{\parallel}(x)$  the {\it physical} component of $f_i(x)$, and $f_i^{\perp}(x)$ the {\it unphysical} one. From (3.14), then, we see that $G^{WW}(x)$ is {\it physical}, and $G^{WN}$ and $G^{NN}(x)$, {\it unphysical}. We note that $\psi^i$ is also decomposed into the {\it physical} component and the {\it unphysical} one by using  $W^{ij}$ and $N^{ij}$ as well as the decomposition of $p_i$.\\
\ts{12pt}Sequentially using the formulas (A.12a) and (A.12b) together with the operator-constraint conditions (3.8), now, we obtain the {\it standard} Hamiltonian $H_S^R$ in the decomposed form as follows:
$$
\begin{array}{lcl}
H_S^R &=& \Pbhat_S H_P\vs{12pt}\\
 &=& H_S^{(\parallel, \parallel)} + H_S^{(\parallel,\perp)} + H_S^{(\perp,\perp)},
\end{array}
\eqno{(5.14a)}
$$
where
$$
\begin{array}{lcl}
H_S^{(\parallel ,\parallel)} &=& \Pbhat_S H_0^{(\parallel ,\parallel)} = \dps{\f12}\pv_i\pv_i + \f{\hbar^2}8G^{WW}(x),\vs{12pt}\\ 
H_S^{(\parallel ,\perp)} &=& \Pbhat_S H_0^{(\parallel ,\perp)} = - \dps{\f{\hbar^2}4}(G^{WW}(x) + G^{WN}(x)),\vs{12pt}\\
H_S^{(\perp ,\perp)} &=& \Pbhat_S H_0^{(\perp ,\perp)} = \left\{\begin{array}{l}
\dps{\f{\hbar^2}8}(G^{WN}(x) + G^{NN}(x))\hs{24pt}(\Pbhat = \Pbhat^I),\vs{12pt}\\
\dps{\f{\hbar^2}8}(2G^{WN}(x) + G^{NN}(x))\hs{24pt}(\Pbhat = \Pbhat^{II})
\end{array}\right.
\end{array}
\eqno{(5.14b)}
$$
with $(x^i , p_i)\in \calC^{(R)}$. The supersymmetric Hamiltonian (4.10a) is rewritten by using the decomposition (5.9) and the projections (5.11) as
$$
H^{SUSY} = \f12\pv_i\pv_i - \f{\hbar^2}8G^{WW}(x),
\eqno{(5.15)}
$$
which is obviously {\it physical}. 
Observing (5.14) and (5.15), then, we obtain the following results:\\
(1) The {\it physical} terms of $H_S^R$, which consist of $\pv_i$ and $G^{WW}(x)$, just reproduce the supersymmetric Hamiltonian (5.15).\\
(2) The discrepancy $\triangle H_S^Q$ is caused by the {\it unphysical} terms consisting of $G^{WN}(x)$ and $G^{NN}(x)$ in $H_S^R$. 

\section{Discussions and Concluding Remarks}

\ts{12pt}We have investigated within the operator formalism of  the constrained systems the quantum mechanics of the $N=1$ supersymmetric dynamical system constrained to the hypersurface $\Sigma^{n-1}$ embedded in the Euclidean space $R^n$ by using the POM. Then, we have obtained the following results:\\
(1) The commutator algebra of the {\it reduced} set $\calC^{(R)}$ is just equivalent to the commutator algebra constructed from the corresponding Dirac brackets.\\
(2) The {\it resulting} Hamiltonian $H_R$ contains the additional terms with the $\hbar^2$-contribution, which are caused by the {\it reduction} of the unconstrained {\it primary} system $(\calC^{(0)}, H_P, \calS^{(0)})$ to the {\it resulting} system $(\calC^{(R)}, H_R, \calS^{(R)})$. The supersymmetric Hamiltonian $H^{SUSY}$ also contains the $\hbar^2$-contribution term in  $(\calC^{(R)}, H_R, \calS^{(R)})$, which is caused by the noncommutativity of the fermionic operators with the bosonic ones. Because of the discrepancy of the additional terms in $H_R$ with the additional term in $H^{SUSY}$, then, the {\it resulting} system does not always preserve the $N=1$ supersymmetry.\\
(3) When the {\it unphysical} additional terms (at least the $G^{NN}$-term) commute with the supercharge $Q^{(R)}$, one can construct the {resulting} system, in which the $H_R$ holds $N=1$ supersymmetry.\\
\ts{12pt}We finally discuss the operators $G^{WW}(x)$, $G^{WN}(x)$ and $G^{NN}(x)$ appearing in the additional terms. Although these operators have appeared in $H_R$ through  the  projections of $H_P$ and in $H^{SUSY}$ through the reordering of the double symmetrized product (4.7), they can be reproduced as follows: Consider the commutators of the normal unit operator $n^i$ with $\pv_i$ and $\pn_i$ in $(\calC^{(0)}, H_P, \calS^{(0)})$. Then, $G^{WW}(x)$ and $G^{WN}(x)$ can be rewritten as follows:
$$
G^{WW} = -\f1{\hbar^2}[n^i, \pv_j][n^i, \pv_j],
\eqno{(6.1)}
$$
$$
G^{WN} = -\f1{\hbar^2}[n^i, \pn_j][n^i, \pn_j].
\eqno{(6.2)}
$$
Using the commutation relations of $\pn_i$ with $n^i$ and $G^i$, which is also normal to $\Sigma^{n-1}$, we obtain
$$
G^{NN}(x) = - \f1{\hbar^2}(g^{-2}[G^i, \pn_j][G^i, \pn_j] - [n^i, \pn_j][n^i, \pn_j]).
\eqno{(6.3)}
$$ 
The relation (6.1) also holds in $(\calC^{(R)}, H_R, \calS^{(R)})$. Thus, $G^{WW}(x)$ is intepreted as to be caused by the quantum {\it correlations} of $n^i$ with $\pv_i$ in $(\calC^{(R)}, H_R, \calS^{(R)})$.  On the other hand, $\pn_i$ vanish by the projection (5.11) in $(\calC^{(R)}, H_R, \calS^{(R)})$. From (6.2) and (6.3), therefore, we find that $G^{WN}(x)$ and $G^{NN}(x)$ are caused by the quantum fluctuations of $G^i$ and $n^i$ due to the uncertainty principle in $(\calC^{(R)}, H_R, \calS^{(R)})$.

\appendix 

\section{Projection Operator Method}

\subsection{Construction of projection operator}

\ts{12pt}Consider second class constraints
$$
T_\alpha({\cal C}) = 0\hs{36pt}(\alpha = 1,\cdots , 2M)
\eqno{(A.1)}
$$
with $\epsilon(T_\alpha) = t$, where ${\cal C} = \{(q^1, p_1), \cdots , (q^N , p_N)\}\ (N > M)$ is a set of {\it initial} canonically conjugate pairs , which has the canonical commutation relations (CCR)
$$
[q^i , p_j] = i\hbar\delta^i_j,\hs{24pt}[q^i, q^j] = [p_i , p_j] = 0
\eqno{(A.2)}
$$
\ts{12pt}The first step is to construct the canonically conjugate set of operators, which we call the associated canonically conjugate set ({\it ACCS}), from the constraint operators $T_\alpha$. Let the {\it ACCS} be $\{(\xi^1, \pi_1),  \cdots , (\xi^M , \pi_M)\}$ with $\epsilon(\xi^a) = \epsilon(\pi_a) = s$\e $(a = 1, \cdots ,M)$. In order to collectively represent the pair $(\xi^a , \pi_a)$, we introduce the $2M$ operators $Z_\alpha$ $(\alpha = 1, \cdots , 2M)$ as follows:
$$
Z_\alpha = \left\{ \begin{array}{l}
\xi^a\hs{12pt}(\alpha = a)\\
\pi_a\hs{12pt}(\alpha = a + M)\hs{24pt}(a = 1, \cdots ,M)
\end{array}\right.
\eqno{(A.3)}
$$
Then, the CCR of $(\xi^a , \pi_a)$ is represented as
$$
[Z_\alpha , Z_\beta] = -i\hbar (-1)^sJ_{\alpha\beta} ,
\eqno{(A.4)}
$$
where $J_{\alpha\beta}$ is the inverse of the $2M\times 2M$ supersymplectic matrix $J^{\alpha\beta}$ defined by
$$
J^{\alpha\beta} = \left(\begin{array}{rr}
0 & I\\
-(-1)^sI & 0
\end{array}\right)_{\alpha\beta}
\eqno{(A.5)}
$$
with the $M\times M$ identity matrix $I$. \\
\ts{12pt}Let $\varphi_\alpha$ the c-number variables $(\epsilon(\varphi) = \epsilon(Z), \alpha = 1, \cdots , 2M)$. Define, then, $\Phat$ by
$$
\Phat = \exp[(-1)^s\Zp_\alpha\frac{\partial}{\partial \varphi_\alpha}]\exp[J^{\alpha\beta}\varphi_\alpha\Zm_\beta]\mid_{\varphi=0}.
\eqno{(A.6)}
$$
From the algebraic relations (1.5) and the commutation relations (A.4), $\Zp_\alpha$ and $\Zm_\alpha$ obey the commutator algebra
$$
[\Zpm_\alpha ,\Zmp_\beta] = -(-1)^sJ_{\alpha\beta},\hs{36pt}[\Zpm_\alpha ,\Zpm_\beta] = 0. 
\eqno{(A.7)}
$$
Using the Baker-Campbell-Hausdorff formula, then, $\Phat$ is shown to have the following properties:
$$
\Phat Z_\alpha = 0,
\eqno{(A.8a)}
$$
$$
\Phat\cdot\Phat = \Phat,
\eqno{(A.8b)}
$$
$$
\Phat\Zp_\alpha = \Zm_\alpha\Phat = 0,
\eqno{(A.8c)}
$$
and satisfy the decomposition of unity,
$$
\exp[-(-1)^s\Zp_\alpha\frac{\partial}{\partial \phi_\alpha}]\Phat\exp[J^{\alpha\beta}\phi_\alpha\Zm_\beta]\mid_{\phi=0} =1,
\eqno{(A.9a)}
$$
$$
\exp[(-1)^s\Zp_\alpha\frac{\partial}{\partial \phi_\alpha}]\Phat\exp[-J^{\alpha\beta}\phi_\alpha\Zm_\beta]\mid_{\phi=0} =1.
\eqno{(A.9b)}
$$
Through the operation of $\Phat$, then, the operators in ${\cal C}$ are transformed as follows:
$$
{\cal C}\ \mapsto \Phat{\cal C} = \{(\Phat q^1, \Phat p_1), \cdots ,(\Phat q^N , \Phat p_N)\},
\eqno{(A.10a)}
$$
$$
Z_\alpha \ \mapsto  \Phat Z_\alpha = 0.
\eqno{(A.10b)}
$$
Because of the noncommutativity, however, the constraints (A.1) do not always hold under the operation of $\Phat$ even though the conditions (A.10b) hold. Therefore, the {\it ACCS} is needed to be constructed in such a manner as the conditions 
$$
\Phat T_\alpha({\cal C}) = T_\alpha(\Phat {\cal C}) = 0
\eqno{(A.11)}
$$
hold, which we call the {\it projection conditions}.

\subsection{Commutator-formula and product-formula}

\ts{12pt}Using (A.7), (A.8) and (A.9), we obtain the following formulas with respect to the commutator $[\Phat A , \Phat B]$ and the symmetrized product $\Phat\{A , B\}$:\\
(a) {\it commutator-formula}
$$
\begin{array}{ll}
[\Phat A , \Phat B] = & \dps{\sum^{\infty}_{n=0}\f{(-1)^{sn}}{(2n)!}(\f{\hbar}{2i})^{2n}}J^{\alpha_1\beta_1}\cdots J^{\alpha_{2n}\beta_{2n}}\vs{16pt}\\
 & \times \Phat [\Zm_{\alpha_1}\cdots \Zm_{\alpha_{2n}}A , \Zm_{\beta_1}\cdots \Zm_{\beta_{2n}}B]\vs{16pt}\\
 & + 2\dps{\sum^\infty_{n=0}\f{(-1)^{s(n+\epsilon(A)+1)}}{(2n+1)!}(\f{\hbar}{2i})^{2n+1}}J^{\alpha_1\beta_1}\cdots J^{\alpha_{2n+1}\beta_{2n+1}}\vs{16pt}\\
& \times \Phat \{\Zm_{\alpha_1}\cdots \Zm_{\alpha_{2n+1}}A , \Zm_{\beta_1}\cdots \Zm_{\beta_{2n+1}}B\} ,
\end{array}\eqno{(A.12a)}
$$
(b) {\it product-formula}
$$
\begin{array}{ll}
\Phat\{ A , B \} = & \dps{\sum^{\infty}_{n=0}\f{(-1)^{sn}}{(2n)!}(\f{\hbar}{2i})^{2n}}J^{\alpha_1\beta_1}\cdots J^{\alpha_{2n}\beta_{2n}}\vs{16pt}\\
& \times \{\Phat\Zm_{\alpha_1}\cdots \Zm_{\alpha_{2n}}A , \Zm_{\beta_1}\cdots \Phat\Zm_{\beta_{2n}}B\}\vs{16pt}\\
 & - \dps{\f12\sum^\infty_{n=0}\f{(-1)^{s(n+\epsilon(A)+1)}}{(2n+1)!}(\f{\hbar}{2i})^{2n+1}}J^{\alpha_1\beta_1}\cdots J^{\alpha_{2n+1}\beta_{2n+1}}\vs{16pt}\\
& \times [\Phat\Zm_{\alpha_1}\cdots \Zm_{\alpha_{2n+1}}A , \Phat\Zm_{\beta_1}\cdots \Zm_{\beta_{2n+1}}B] .
\end{array}\eqno{(A.12b)}
$$
\ts{12pt}Successively using the formulas (A.12a) and (A.12b), then, the commutator between $\Phat A$ and $\Phat B$ is represented in the form of the power series of $\hbar$ like
$$
[\Phat A , \Phat B] = \sum^\infty_{n=0}\hbar^{n}C^{(n)}(\Phat{\cal C}),
\eqno{(A.13)}
$$
the first two terms of which are just equivalent to the quantized form of the corressponding Dirc bracket, and the others of which are the quantum corrections caused by the noncommutativity of $Z_\alpha$ with the operators $A$ and $B$. Similarly, $\Phat\{A , B\}$ is represented in such a form as
$$
\Phat\{A , B\} = \sum^\infty_{n=0}\hbar^{2n}S^{(n)}(\Phat{\cal C}) ,
\eqno{(A.14)}
$$
the terms containing $\hbar$ of which give the quantum corrections for the symmetrized product of operators.

\subsection{Fermionic constraints}

\ts{12pt}Consider the other type of {\it ACCS} $z_\alpha({\cal C})$ $(\alpha = 1, \cdots , 2M)$ obeying the CCR
$$
[z_\alpha , z_\beta] = \delta_{\alpha\beta},
\eqno{(A.15)}
$$
which occurs in the case of the Grassmann-odd constraints. From (A.15), $\hat{z}^{(\pm)}$ obey the commutator algebra
$$
[\hat{z}^{(\pm)}_\alpha , \hat{z}^{(\mp)}_\beta] = \f1{i\hbar}\delta{\alpha\beta},\hs{36pt}[\hat{z}^{(\pm)}_\alpha , \hat{z}^{(\pm)}_\beta] = 0.
\eqno{(A.16)}
$$
Comparing the algebra(A.7) with (A.16), thus, one finds that the supersymplectic matrix $J_{\alpha\beta}$ is replaced with the diagonal matrix
$$
{\cal J}_{\alpha\beta} = -(-1)^si\hbar\delta_{\alpha\beta}
\eqno{(A.17)}
$$
in the algebraic formulas containing $J_{\alpha\beta}$. 

\subsection{Successive operation of projection operators}

\ts{12pt}Consider the case that a set of constraint operators is classified into several subsets (for example, $L$ subsets) as follows:
$$
\calS^{(k)} = \{T_\alpha^{(k)}(\calC^{(0)})\mid \alpha = 1,\cdots ,M_k\},\hs{24pt}(k = 1,\cdots , L)
\eqno{(A.18)}
$$
where $\calC^{(0)}$ is a set of the {\it initial} unconstrained canonically
 conjugate pairs. Here, although the {\it ACCS} of each subset $\calS^{(k)}$ should be  assumed to satisfy (A.4), the {\it ACCS} involved in the different subsets would not always be commutable with each other. Let $\calZ^{(k)} = \{Z^{(k)}_1, \cdots ,Z^{(k)}_{M_k}\}$ be a set of the {\it ACCS} associated with $\calS^{(k)}$, and let $\Phat^{(k)}$ be the projection operator constructed with $\calZ^{(k)}$. Consider, then, the successive operations of $\Phat^{(k)}\e (k = 1,\cdots ,L)$ on any operator $O$ with certain ordering of them,
$$
\Phat^{(k_L)}\cdots \Phat^{(k_1)}O,
\eqno{(A.19)}
$$
where $(k_1,\cdots ,k_L)$ denotes certain reordering of $(1,\cdots ,L)$. In (A.19), $\Phat^{(k_1)}$ is constructed with $\calZ^{(k_1)}$ in $\calC^{(0)}$, which we express as
$$
\Phat^{(k_1)} = \Phat^{(k_1)}(\calC^{(0)}),
\eqno{(A.20)}
$$
and thus, $\Phat^{(k_n)}\e (n = 1,\cdots, L)$ are expressed as
$$
\Phat^{(k_n)} = \Phat^{(k_n)}(\calC^{(n-1)})
\eqno{(A.21)}
$$
with
$$
\calC^{(n)} = \Phat^{(k_n)}\calC^{(n-1)}.
\eqno{(A.22)}
$$
Then, $\Phat^{(k)}\e (k=1, \cdots ,L)$ are not always commutable with each other when operating on $O$:
$$
\Phat^{(k)}\Phat^{(l)}O \neq \Phat^{(l)}\Phat^{(k)}O.
\eqno{(A.23)}
$$    
 
\section{Consistency Conditions of Primary Constraints}

\ts{12pt}We here present the secondary constraints required from the consistency conditions for the primary constraints and determine the Lagrange multiplier operators.\\
(1) From the conditions $\chid_i =(1/i\hbar)[\chi_i , H_P] = 0$, the Lagrange multiplier operators $\mu^i$ are determined as
$$
\mu^i = \lambda G_i(x).
\eqno{(B.1)}
$$ 
(2) The condition $\Pid_\lambda = (1/i\hbar)[\Pi_\lambda , H_P] = 0$ requires the sequential secondary constraints
$$
\eta _1 = \psi^iG_i(x) = 0,
\eqno{(B.2a)}
$$
$$
\eta_2 = \{\psi^iG^j_i(x) , p_j\} + \lambda g^2 = 0,
\eqno{(B.2b)}
$$
and determines the Lagrange multiplier operator $\tau$ as
$$
\tau = \{g^{-2} , \tau_0\} + \f14[g^2 , [g^{-2}, \{g^{-2} , \tau_0\}]],
\eqno{(B.2c)}
$$
where
$$
\tau_0 = -\{\{\psi^iG^{jk}_i(x), p_j\}, p_k\} - 3\{\lambda G^i(x)G^j_i(x), p_j\} - X\psi^iG^j_i(x)G_j(x).
\eqno{(B.2d)}
$$ 
(3) The condition $\PXd = (1/i\hbar)[\PXd , H_P] = 0$ requires the sequential secondary constraints
$$
h_1 = G(x) = 0,
\eqno{(B.3a)}
$$
$$
h_2 = \{G^i(x) , p_i\} = 0,
\eqno{(B.3b)}
$$
$$
h_3 = \{\{G^{ij}, p_i\}, p_j\} + i\lambda\psi^iG_{ij}(x)G^j(x) = 0,
\eqno{(B.3c)}
$$
and determines the Lagrange multiplier operator $u$ as
$$
u = \{g^{-2} , u_0\} + \f14[g^2, [g^{-2}, \{g^{-2}, u_0\}]],
\eqno{(B.3d)}
$$
where 
$$
\begin{array}{ll}
u_0 =& -\{\{\{G^{ijk}(x) , p_i\}, p_j\}, p_k\} - 4\{XG_i(x)G^{ij}(x), p_j\} \vs{8pt}\\
&- i\{\lambda\psi^i(3G_{ij}(x)G^{jk}(x)+G^j(x)G_{ijk}(x)), p_k\} - i\{\tau , \psi^iG_{ij}(x)G^j(x)\}.
\end{array}
\eqno{(B.3e)}
$$

\end{document}